# Role of La doping for Topological Hall Effect in Epitaxial EuO Films

Yu Yun[1,2,†], Yang Ma[1,2,†], Tang Su[1,2], Wenyu Xing[1,2], Yangyang Chen[1,2], Yunyan Yao[1,2], Ranran Cai[1,2], Wei Yuan[1,2], and Wei Han[1,2,*]

[1] International Center for Quantum Materials, School of Physics, Peking University, Beijing 100871, P. R. China

[2] Collaborative Innovation Center of Quantum Matter, Beijing 100871, P. R. China

[†] These authors contributed equally to the work

[*] Correspondence to: weihan@pku.edu.cn

**Abstract**

We report the critical role of La doping in the topological Hall effect observed in $La_xEu_{1-x}O$ thin films (~ 50 nm) grown by molecular beam epitaxy. When the La doping exceeds 0.036, topological Hall effect emerges, which we attribute to the formation of magnetic skyrmions. Besides, the La doping is found to play a critical role in determining the phases, densities, and sizes of the skyrmions in the $La_xEu_{1-x}O$ thin films. The maximum region of the skyrmion phase diagram is observed on the $La_{0.1}Eu_{0.9}O$ thin film. As the La doping increases, the skyrmion density increases while the skyrmion size decreases. Our findings demonstrate the important role of La doping for the skyrmions in EuO films, which could be important for future studies of magnetic skyrmions in Heisenberg ferromagnets.

# I. INTRODUCTION

Magnetic skyrmion, a topological particle-like nontrivial spin texture, has attracted a great deal of attention arising from its exotic physical properties and potential application for high density memory devices [1-3]. Dzyaloshinskii–Moriya interaction (DMI), a chiral interaction due to the inversion symmetry breaking, has been demonstrated to be essentially important for skyrmions in bulk non-centrosymmetric B20 chiral magnets [2,4,5], including MnSi [6-8], $Fe_{1-x}Co_xSi$ [9], FeGe [10,11], etc, and their analog $Cu_2OSeO_3$, a multiferroic insulator [12,13]. Besides, interfacial DMI between a ferromagnetic (FM) layer and a heavy metallic (HM) layer with large spin-orbit coupling gives rise to the formation of skyrmions at the FM-HM interface [14-20]. Beyond the DMI, frustrated exchange interaction has been proposed for the formation of the skyrmions in centrosymmetric cubic $SrFe_{1-x}Co_xO_3$ systems [21-23]. Very interestingly, a recent observation of the topological Hall effect (THE) in $EuO_{1-x}$ thin films provides a strong evidence for the presence of skyrmions in classical 2D Heisenberg ferromagnet [24-26]. The non-monotonic thickness dependence of the skyrmion is reported, which exhibit different behaviors compared to the B20 chiral magnets [10,24]. For EuO thin films, it is well known that the ferromagnetic properties can be tuned by carriers in the 5d band, such as rare earth atoms doping and oxygen vacancies [27-33]. However, the role of the carrier concentration and rare earth atoms doping for the skyrmion in EuO films has not been reported yet.

In this paper, we report the critical role of the La doping for the THE observed in $La_xEu_{1-x}O$ thin films, with $x$ systematically varying from 0 to 0.20. The Curie temperature ($T_C$) could be enhanced up to ~ 127 K with a La doping of $x = 0.036$, and exhibits a similar trend with the carrier concentration when the La doping changes. THE, the indirect experimental signature of magnetic skyrmion formation, emerges as the temperature increases close to $T_C$ of $La_xEu_{1-x}O$ thin films, as $x$ ranges from 0.036 to 0.20. Following the established analysis that THE arises from the local effective magnetic field experienced by itinerant electrons, we systematically investigate the role of La doping on the skyrmion density and size from the topological Hall resistivity. As the La doping increases, the skyrmion density increases, and skyrmion size decreases. These results present the critical role of La doping in modulating the formation of skyrmions, skyrmions phases, skyrmion densities, and skyrmion sizes in Heisenberg ferromagnetic $La_xEu_{1-x}O$ thin films.

## II. METHODS

Epitaxial $La_xEu_{1-x}O$ thin films are grown on (001)-oriented yttrium-stabilized cubic zirconia (YSZ) substrates using oxide molecular beam epitaxy (MBE-Komponenten GmbH; Octoplus 400). Prior to the film growth, the YSZ substrates are annealed for 1 hour at 600 ºC and then cooled down to 450 ˚C. For EuO thin films, a very thin layer of EuO (~ 6 nm) is grown using thermally evaporation of Eu with a deposition rate of ~ 8 Å/min [34]. Then oxygen is introduced and its partial pressure is maintained at $1.5 \times 10^{-9}$ mbar to proceed the growth of EuO. For $La_xEu_{1-x}O$ thin films, La and Eu are co-evaporated from thermal effusion cells with different La deposition rates to control the atomic ratio. During growth, *in situ* reflective high energy electron diffraction (RHEED) is used to monitor and characterize the films' crystalline quality. At the end, a thin MgO layer (~ 5 nm) is deposited *in situ* via e-beam deposition as a capping layer to avoid sample degradation with air exposure.

The crystalline structure is also characterized using high-resolution X-ray diffraction (XRD) with a Bruker D8 Discover system. The ferromagnetic properties are studied using a Magnetic Properties Measurement System (MPMS; Quantum Design). The topological Hall effect and electron transport properties are measured using either Van der Pauw technique or standard Hall bar configuration in an Oxford Spectromag system.

**III. RESULTS**

The RHEED patterns of the YSZ (001) substrates and the epitaxial La$_x$Eu$_{1-x}$O films (~ 50 nm) viewed from both [100] and [110] directions are shown in Fig. 1(a-h). Sharp RHEED patterns are observed for all the La$_x$Eu$_{1-x}$O films with $x$ from 0 up to 0.20. The RHEED patterns of La$_x$Eu$_{1-x}$O films show narrow full width of half maximums (FWHMs) (~ 2.3 - 6.2 pixels for [100] direction and ~ 2.0 – 8.0 pixels for [110] direction), which are similar to those of YSZ substrates (~ 4.8 pixels for [100] direction and ~ 2.3 pixels for [110] direction). Fig. 2(a) shows the XRD results for three typical La$_x$Eu$_{1-x}$O films. Clear Laue fringes around the (002) peak of EuO further indicate the highly crystalline quality, and from which we estimate the film thickness to be ~ 50 nm. A slight shift of the (002) peak of the epitaxial La$_x$Eu$_{1-x}$O films is observed as the La doping increases. The c-axis lattice constant is found to exhibit a tiny increase (< 0.6%) with La doping, as shown in Fig. 2(b).

The magnetic moments of the La$_x$Eu$_{1-x}$O films are measured as a function of temperature with an in-plane magnetic field of 1000 Oe. The onset of the magnetic moment could be used to identify T$_C$, as shown in Fig. 2(c). As the La doping increases, long range ferromagnetic order strengthens and a maximum T$_C$ of ~ 127 K is observed on the La$_{0.036}$Eu$_{0.964}$O film. The double-dome feature of temperature-dependent magnetization curves for La$_x$Eu$_{1-x}$O ($x > 0$) films is attributed to the Ruderman-Kittel-Kasuya-Yosida (RKKY) interactions in the presence of conducting carriers [33]. As the La doping further increases, the T$_C$ decreases. The La doping dependence of T$_C$ is summarized in Fig. 2(d). To understand this phenomenon, the carrier densities of the films are measured using Hall effect in a perpendicular magnetic field from -5 T to 5 T (See supplementary information S1). The carrier density at 2K also exhibits a peak at the La doping of $x = 0.036$, which is exactly the same doping level for the film with highest $T_C$. The similar La doping dependences of the $T_C$ and the carrier density are consistent with previous study on Gd doped EuO films [27].

A typical magnetic field dependence of the Hall resistivity is shown in Fig. 3(a), measured on the 50 nm $La_{0.10}Eu_{0.90}O$ thin film at 70 K. Apart from the linear Hall region where the magnetic field is larger than 3 T, a strong non-linear Hall effect is observed, which is associated with the magnetic properties of the film. After subtracting linear ordinary Hall contribution, the residual Hall resistivity, the sum of the anomalous hall and topological Hall resistivities, is shown in Fig. 3(b). The red and black lines indicate the results with magnetic fields sweeping up and down, respectively, between -4 T and 4 T. The saturation of the Hall resistivity at high field is consistent with the anomalous Hall effect, that is the full alignment of the magnetization above the saturation field (blue lines in Fig. 3(b)). The temperature dependences of the anomalous Hall resistivity ($\rho_{AH}$) and the longitudinal resistivity of the films are investigated, presenting a relationship of $\sigma_{AH} \propto \sigma_{xx}^{2.5}$, where $\sigma_{AH}$ and $\sigma_{xx}$ are the anomalous Hall and the longitudinal conductivities (See supplementary information S2). Besides the conventional anomalous Hall signal, the topological Hall resistivity is also observed, which is attributed to the formation of the skyrmions under such magnetic fields. As the anomalous Hall component scales with the magnetization ($R_{AH} = R_S M$), the topological Hall resistivity could be obtained by subtracting the product of magnetization and a constant, as shown in Fig. 3(c). Given the low conductivities for $La_xEu_{1-x}O$ films ($< 3 \times 10^3\ S \cdot cm^{-1}$), the topological Hall resistivity observed here cannot be attributed to skew scattering [35]. In such dirty limit, the intrinsic contribution from the Berry curvature is dominant, in contrast to the clean limit where the skew scattering is dominant [36,37]. The THE results measured at various temperatures are shown in Fig. 3(d).

To probe the mechanism for the formation of skyrmions in these ~ 50 nm $La_xEu_{1-x}O$ thin films, THE is studied as a function of the temperature and our-of-plane magnetic field for a series of $La_xEu_{1-x}O$ films with $x$ systematically increasing from 0.005 to 0.20. When the La doping is relatively low ($x$ = 0.005, 0.009), an unconventional Hall resistivity is observed at very low temperatures (See supplementary information S3). These unconventional Hall resistivity is different from THE due to skyrmions, which are usually observed at the elevated temperatures slightly below $T_C$ [1,2,6-11]. The mechanisms of these unconventional AHE in low doped $La_xEu_{1-x}O$ thin films need further studies. When the La doping exceeds 0.036, a topological Hall effect is observed on the ~50 nm $La_xEu_{1-x}O$ thin films with $x$ = 0.036, 0.10, 0.15, and 0.20, as shown in Fig. 4(a-d). For the $La_xEu_{1-x}O$ thin films with $x$ = 0.036, skyrmions emerge at the temperatures above ~50 K (Fig. 4(a)). With the increase of La dopant concentration ($La_{0.10}Eu_{0.90}O$), the skyrmions stability extends to a much wider temperature scope, as shown in Fig. 4(b). As the La doping further increases, the skyrmion phase is suppressed, indicated in Fig. 4(c-d) for the $La_{0.15}Eu_{0.85}O$ and $La_{0.20}Eu_{0.80}O$ thin films respectively. We note that the maximum region for the presence of skyrmions is observed on the ~ 50 nm $La_{0.10}Eu_{0.90}O$ thin film.

**IV. DISCUSSION**

These results indicate that the doping level of La is extremely important for the skyrmion phase diagram and the amplitude of the topological Hall resistivity. To understand the mechanism of the La doping modulation effect of the topological Hall resistivity in these $La_xEu_{1-x}O$ films, the skyrmion densities and sizes are estimated. It has been demonstrated in previous studies that the topological Hall resistivity stems from the extra transverse velocity of electrons when they feel the locally effective magnetic fields exerted by the skyrmions [7,11,38-40]. Thus, a larger topological Hall resistivity is expected with a higher density of the skyrmions. The skyrmions density and sizes are estimated using the following relationship [7,11,38-40]:

$$\rho_{TH} = PR_H n_{sk} \phi_0 \tag{1}$$

where $P$ is the spin polarization, $R_H$ is the ordinary Hall coefficient, $n_{sk}$ is the skyrmions density, and $\phi_0$ is one flux quantum. The skyrmion size is estimated using the length scale of a single skyrmion $n_{sk}^{-1/2}$ and $P = 80\%$ from previous study [28]. As shown in Fig. 5(a), the calculated skyrmion size decreases as the La doping increases. The skyrmion density is found to monotonically increase as the La doping increases, as shown in Fig. 5(b). These results demonstrate the critical role of the La doping for skyrmions observed in the $La_xEu_{1-x}O$ thin films.

Regarding the underlying causes of the skyrmions observed in the $La_xEu_{1-x}O$ thin films, there are several possible mechanisms. The first possible reason is related to the long range magnetic dipolar interaction, which has been shown to give rise to the formation of skyrmions from ~ 100 nm to 1 μm, sharing a similar length scale with skyrmions observed here [1]. In this scenario, the essential reason is the competition between the dipolar interaction and out-of-plane anisotropy. However, the $La_xEu_{1-x}O$ thin films show in-plane anisotropy, and the long range magnetic dipolar interaction seems unlikely. The second possible reason is related to the induced carriers, which can enhance the $T_C$ due to the RKKY coupling of these carriers. A recent theoretical calculation shows the stabilization of magnetic skyrmions due to RKKY interactions [41]. However, our results show that the maximum $T_C$ is observed for the La$_{0.036}$Eu$_{0.964}$ film, while the maximum phase diagram of skyrmions is observed on the La$_{0.10}$Eu$_{0.90}$ film. The third one is related to the magnetic frustration with competing ferromagnetic and antiferromagnetic interactions [42]. This mechanism could explain the observation of strongest THE around the $T_C$, below which the spins of the $f$ electrons on Eu atoms favor ferromagnetic alignment, while above which the spins of the $f$ electrons favor antiferromagnetic alignment. But it fails to explain the observation of THE at lower temperatures far below $T_C$. The last reason is associated with La$^{3+}$ impurities. The extrinsic La$^{3+}$ impurities generate local electric fields, and we speculate that these impurities will break the inversion symmetry and might give rise to local DMI with a vector direction perpendicular to the plane of La and its nearest Eu atoms. Following this scenario, a higher doping of the La$^{3+}$ impurities gives rise to a higher density of skyrmions, which agrees with our experimental observations. Nevertheless, to fully understand physical mechanisms of the skyrmions observed in the La$_x$Eu$_{1-x}$O thin films, future theoretical studies are needed.

To be noted, these THE experimental results are indirect experimental evidences of the skyrmions in La$_x$Eu$_{1-x}$O thin films. The direct imaging techniques, including Lorentz transmission electron microscopy, nitrogen-vacancy quantum sensing, spin-polarized low-energy electron microscopy, and X-ray microscopy, would be essential to further confirm the formation of skyrmions, which might lead to unusual characteristics in magnetic skyrmions in Heisenberg ferromagnets [9,43-45].

**V. CONCLUSION**

In summary, the important role of the La doping for the topological Hall effect in $La_xEu_{1-x}O$ thin films is reported. The $T_C$ and carrier concentration exhibit similar trends as a function of the La doping. THE is observed in $La_xEu_{1-x}O$ thin films with *x* varying from 0.036 to 0.20, and the maximum range of the skyrmion phase is found in the $La_{0.10}Eu_{0.90}O$ film, where the THE is observed down to the temperature of 2 K. Skyrmion size shrinks continuously with increasing La doping, and skyrmion density increases with higher La concentration. Our results demonstrate the route of extrinsic doping in the modulation of the skyrmions. These experimental results could be important for future theoretical understandings and experimental searches of magnetic skyrmions in Heisenberg ferromagnets.

**ACKNOWLEDGEMENTS**

We acknowledge the fruitful discussion with Jia Li and Ryuichi Shindou. We also acknowledge the financial support from National Basic Research Programs of China (973 program Grant Nos. 2015CB921104 and 2014CB920902) and National Natural Science Foundation of China (NSFC Grant No. 11574006 and 11704011).

**Figure 1**

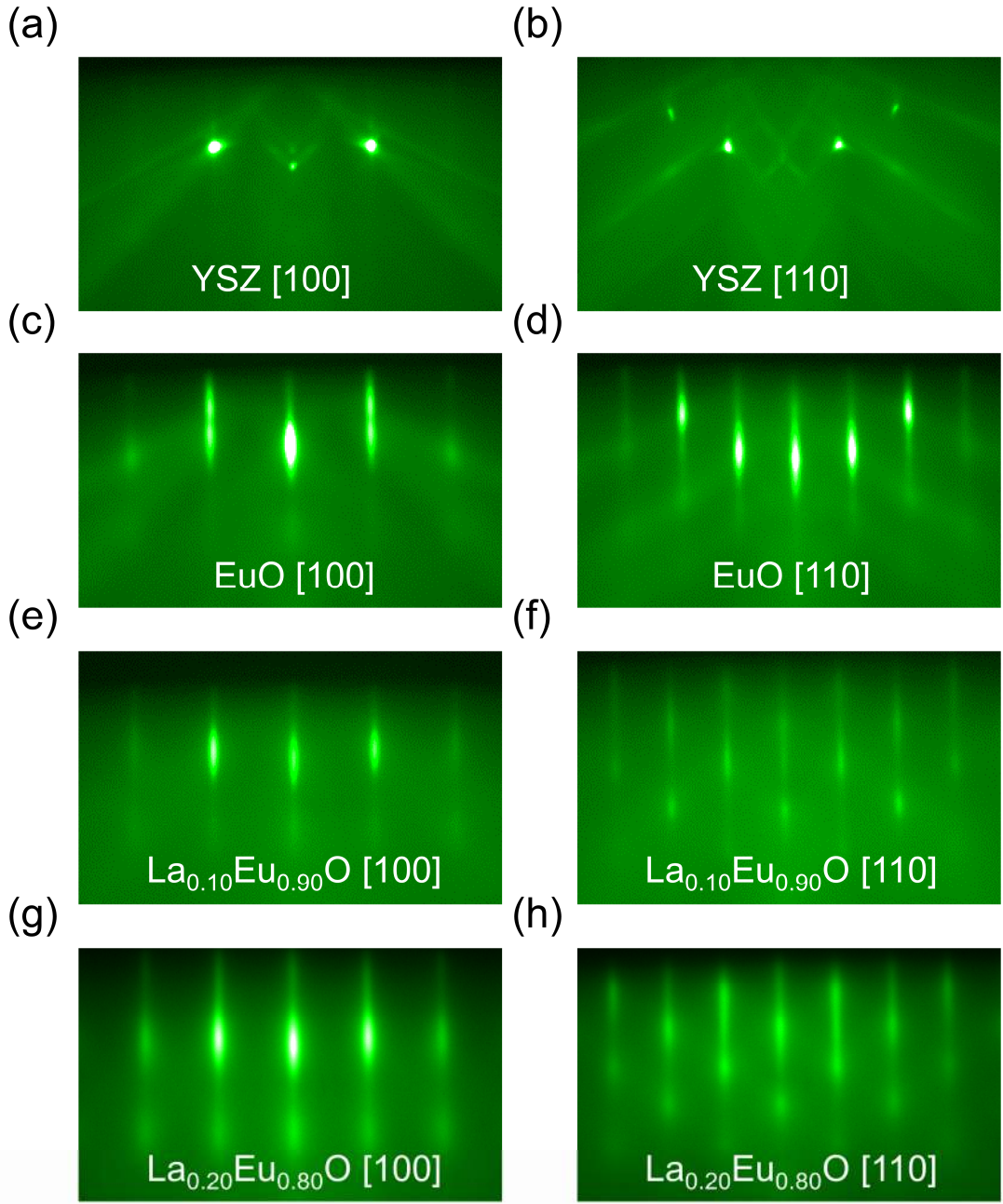

FIG. 1. *In situ* RHEED characterization for the YSZ substrates and $La_xEu_{1-x}O$ thin films. (a-d) RHEED patterns of the YSZ substrates and EuO thin films viewed along the [100] and [110] directions. (e-h) RHEED patterns of the $La_xEu_{1-x}O$ thin films with $x$ = 0.10, and 0.20, respectively.

**Figure 2**

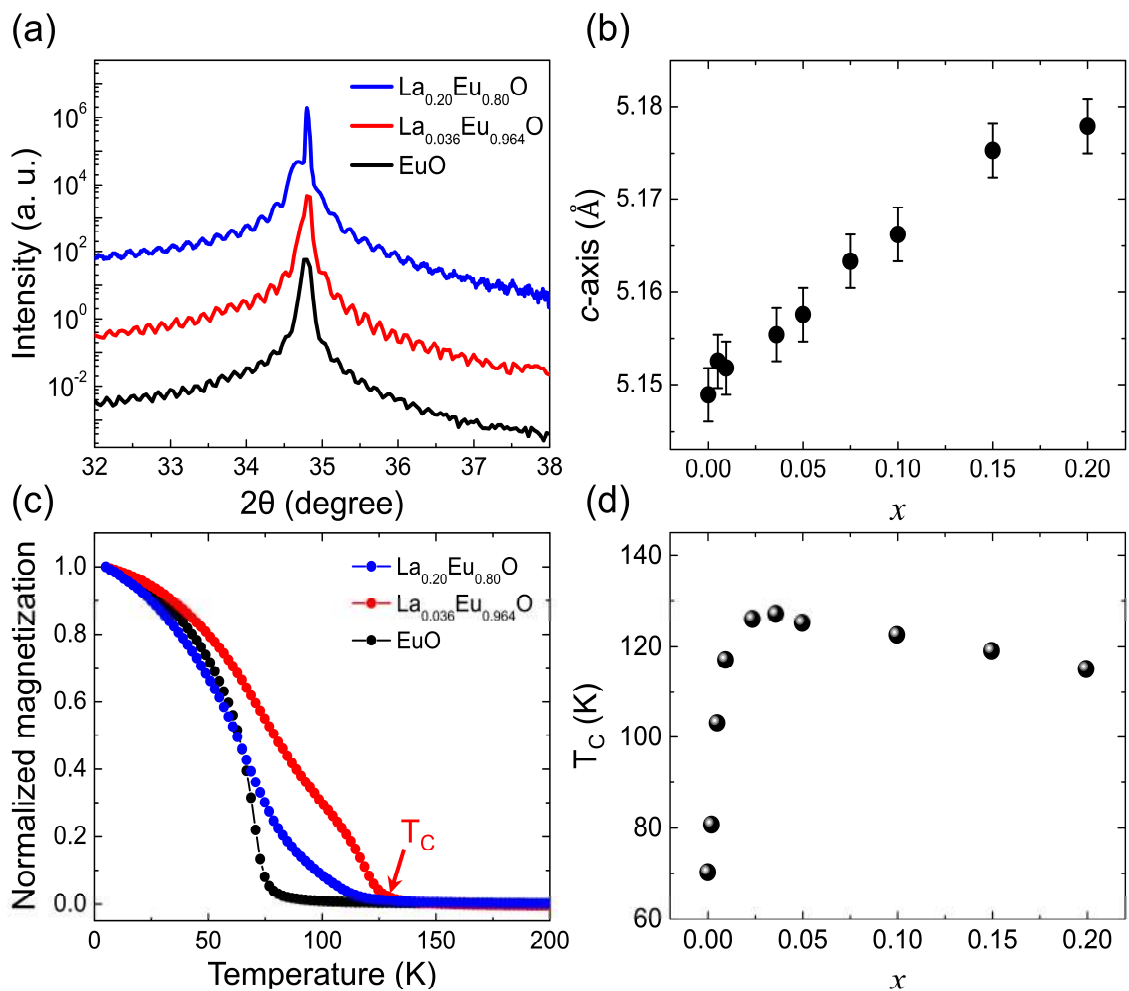

FIG. 2. The crystal structures and ferromagnetic properties of the La$_x$Eu$_{1-x}$O thin films as a function of the La doping. (a) XRD results of the La$_x$Eu$_{1-x}$O thin films (~ 50 nm). Laue fringe peaks are observed around the (002) main peak of EuO. (b) The *c*-axis lattice constant as a function of La doping. (c) The temperature dependence of the normalized magnetization measured on these La$_x$Eu$_{1-x}$O thin films with an in-plane magnetic field of 1000 Oe. The onset of the magnetization is used to identify the Curie temperature, as indicated by the arrow. (d) The T$_C$ of the La$_x$Eu$_{1-x}$O thin films as a function of La doping.

**Figure 3**

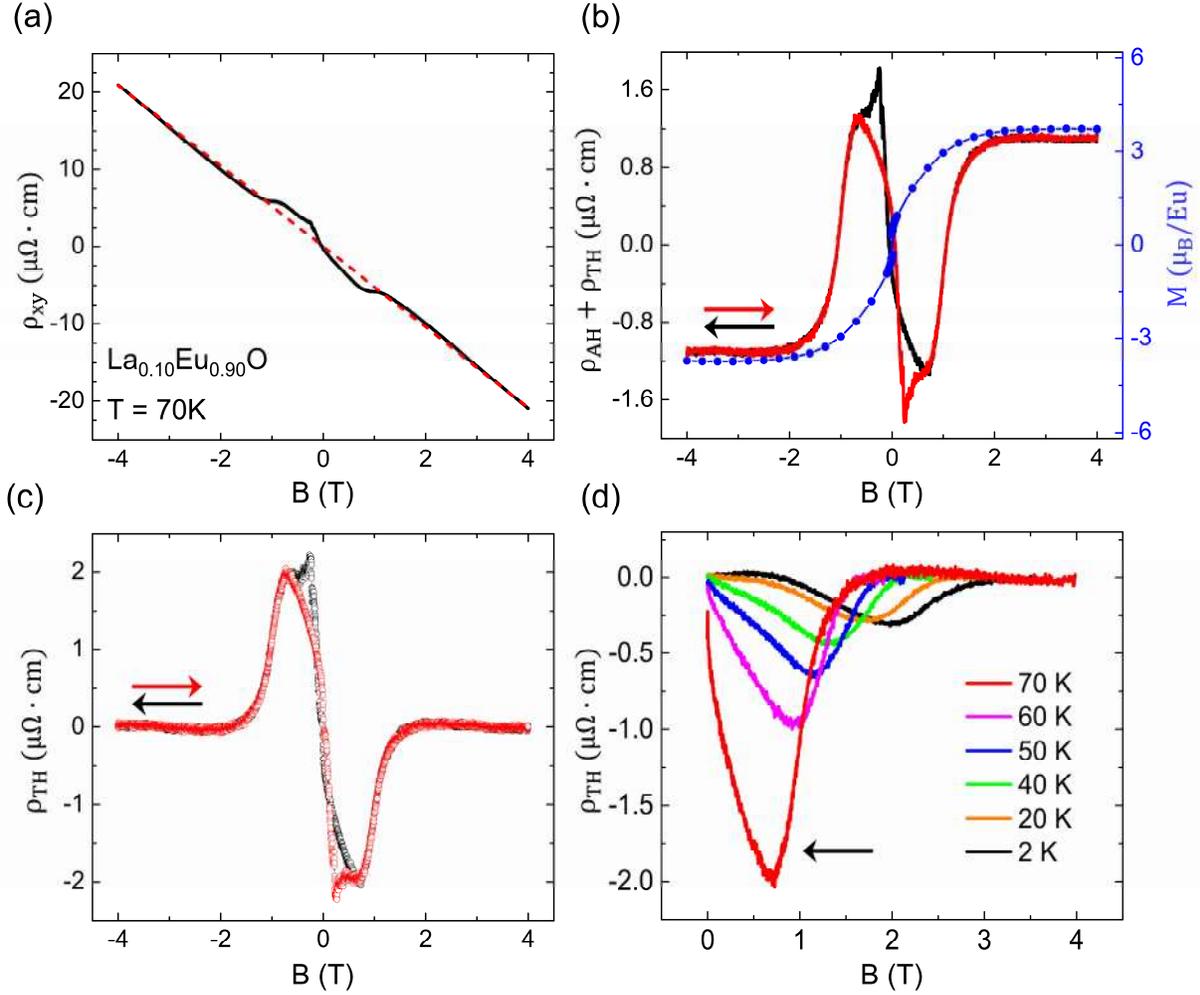

FIG. 3. The Hall measurement on a typical film (~ 50 nm $La_{0.1}Eu_{0.9}O$). (a) Magnetic field dependence of Hall resistivity at 70 K. (b) The sum of anomalous and topological Hall resistivities ($\rho_{AH} + \rho_{TH}$) at 70K as a function of the out-of-plane magnetic field after subtraction of the ordinary Hall contribution. The blue curve is the magnetic field dependence of the magnetization. (c) The topological Hall resistivity ($\rho_{TH}$) at 70K as a function of the out-of-plane magnetic field. (d) The topological Hall resistivity ($\rho_{TH}$) as a function of the out-of-plane magnetic field measured at different temperatures.

**Figure 4**

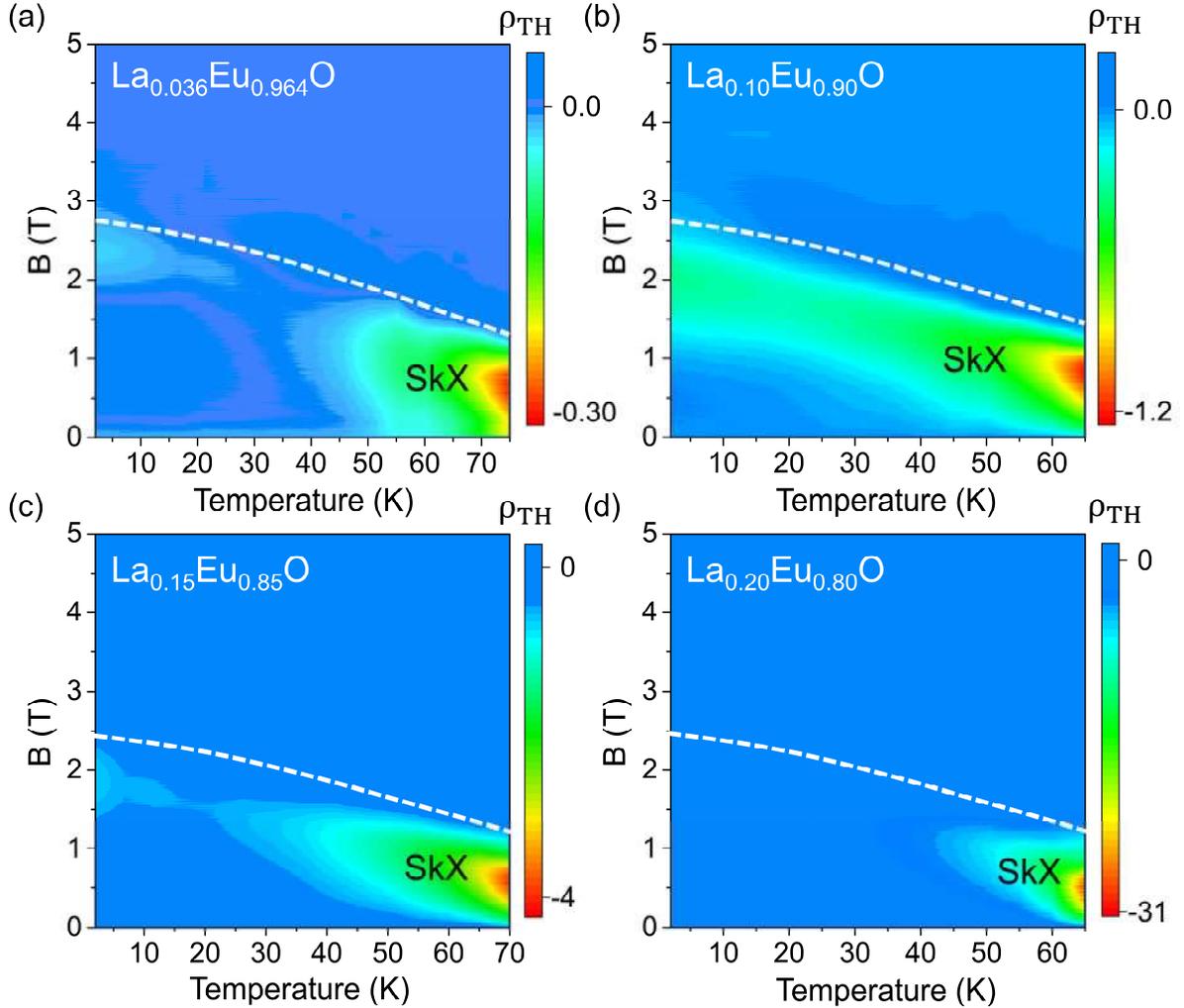

FIG. 4. The phase diagrams of topological Hall resistivity $\rho_{TH}$ ($\mu\Omega \cdot cm$) in the $La_xEu_{1-x}O$ thin films as a function of magnetic field and temperature. (a-d) The topological Hall resistivity as a function of the magnetic field and temperature measured on the $La_{0.036}Eu_{0.964}O$, $La_{0.10}Eu_{0.90}O$, $La_{0.15}Eu_{0.85}O$, and $La_{0.20}Eu_{0.80}O$ thin films, respectively. The dashed white lines are guide lines for the saturation magnetic field as a function of the temperature. The skyrmions regions are marked with "SkX".

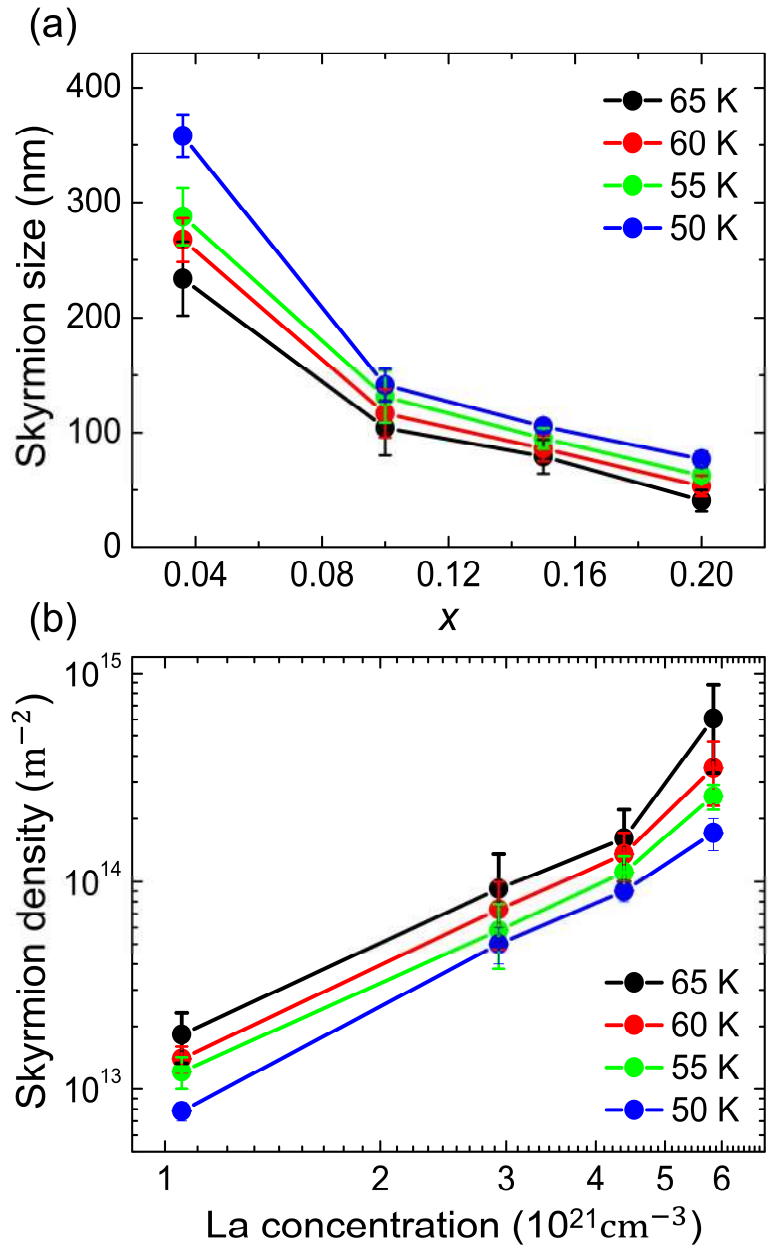

FIG. 5. The important role of La doping for the skyrmion density and size. (a) The estimated skyrmion size as a function of La doping *x*. (b) La concentration dependence of skyrmion density.

# Supplementary Information:

# Role of La doping for Topological Hall Effect in Epitaxial EuO Films

Yu Yun[1,2,†], Yang Ma[1,2,†], Tang Su[1,2], Wenyu Xing[1,2], Yangyang Chen[1,2], Yunyan Yao[1,2], Ranran Cai[1,2], Wei Yuan[1,2], and Wei Han[1,2,*]

[1] International Center for Quantum Materials, School of Physics, Peking University, Beijing 100871, P. R. China

[2] Collaborative Innovation Center of Quantum Matter, Beijing 100871, P. R. China

[†] These authors contributed equally to the work

[*] Correspondence to: weihan@pku.edu.cn

## S1. Characterization of the carrier density

To determine the carrier density, Hall measurements are performed in an Oxford Spectromag system. Two different geometries, Van der Pauw on unpatterned films and Hall bar devices via standard photolithography and etching techniques, are used in our experiments. It has been confirmed that similar results are obtained on the same the La$_x$Eu$_{1-x}$O film with both measurement geometries. During the measurement, transverse resistivity ($\rho_{xy}$) is recorded when the magnetic field is swept between -5 T and 5 T. The carrier density is calculated based on the formula of $n_{3D} = -\frac{1}{R_H \cdot e \cdot t}$, where R$_H$ is the slope of the resistivity as a function of the magnetic field, $e$ is the electron charge, and $t$ is the film thickness.

The carrier densities obtained on these La$_x$Eu$_{1-x}$O thin films are shown in Fig. S1(a). As the temperature changes, the carrier density exhibit little variation. It is noted that the carrier density first increases, and then decreases as the La doping increases. Fig. S1(b) shows La doping dependence of the carrier density at 2K. The maximum of the carrier density is observed on the La$_{0.036}$Eu$_{0.964}$O thin film, which also shows the maximum of the Curie temperature (T$_C$). As La doping increases from 0.036 to 0.20, both the T$_C$ and the carrier density deceases, suggesting that a large part of extrinsic La atoms remains inactive. Overall, the similar La doping dependences of the T$_C$ and the carrier density present the importance of the enhancement of the T$_C$ due to Ruderman-Kittel-Kasuya-Yosida (RKKY) interactions from the conducting carriers.

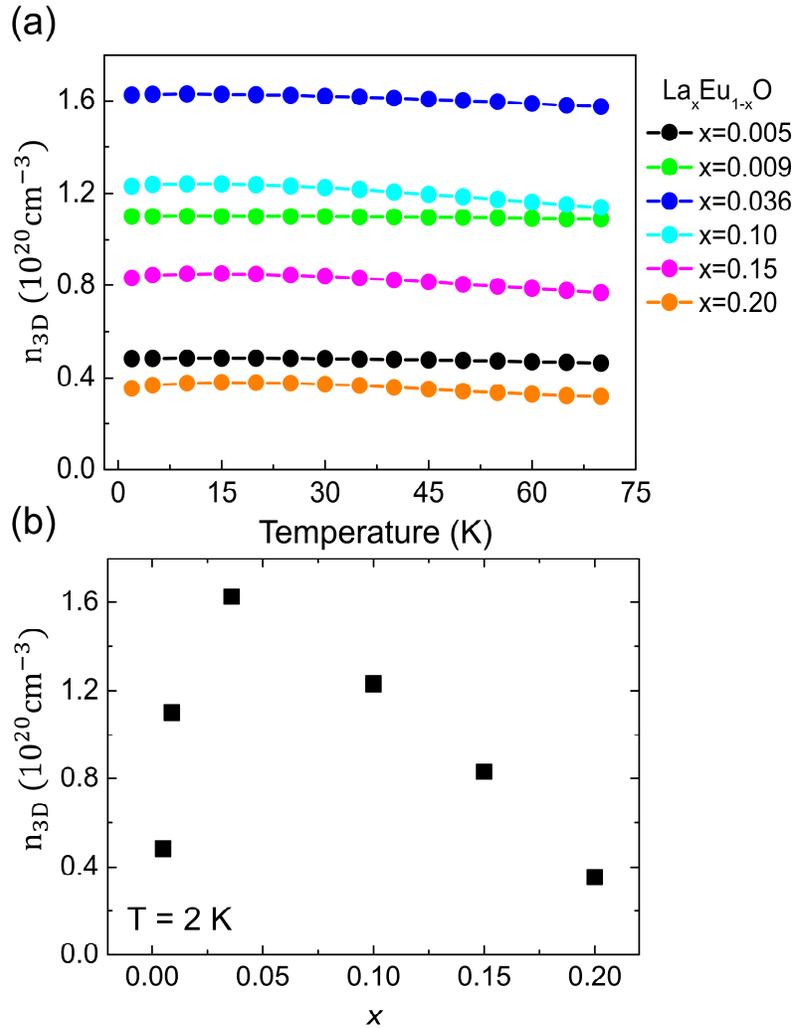

FIG. S1. Temperature and La doping dependences of the carrier density for $La_xEu_{1-x}O$ thin films. (a) The carrier density as a function of temperature for $La_xEu_{1-x}O$ thin films with $x$ varying from 0.005 to 0.20. (b) The carrier density at 2 K as a function of La doping.

## S2. Anomalous Hall effect in $La_{0.10}Eu_{0.90}O$ thin film

The relationship of the anomalous Hall resistivity ($\rho_{AH}$) and longitudinal resistivity ($\rho_{xx}$) $La_{0.10}Eu_{0.90}O$ thin film is investigated via varying the temperature. As the temperature decreases below the $T_C$, the longitudinal resistivity decreases and saturates at low temperatures, as indicated by black circles in Fig. S2(a). Whileas, the anomalous Hall resistivity exhibits a non-monotonic variation (red circles in Fig. S2(a)). The anomalous Hall conductivity ($\sigma_{AH}$) is plotted as a function of the longitudinal conductivity ($\sigma_{xx}$) (Fig. S2(b)). A relationship of $\sigma_{AH} \propto \sigma_{xx}^{2.5}$ is identified for the higher temperature region, as indicated by the blue dashed in line in Fig. S2b. However, the results for the lower temperature region exhibits different behavior, indicating the complexity of the anomalous Hall effect in La$_x$Eu$_{1-x}$O. Future investigations are needed to identify the mechanisms for the anomalous Hall effect in La$_x$Eu$_{1-x}$O thin films.

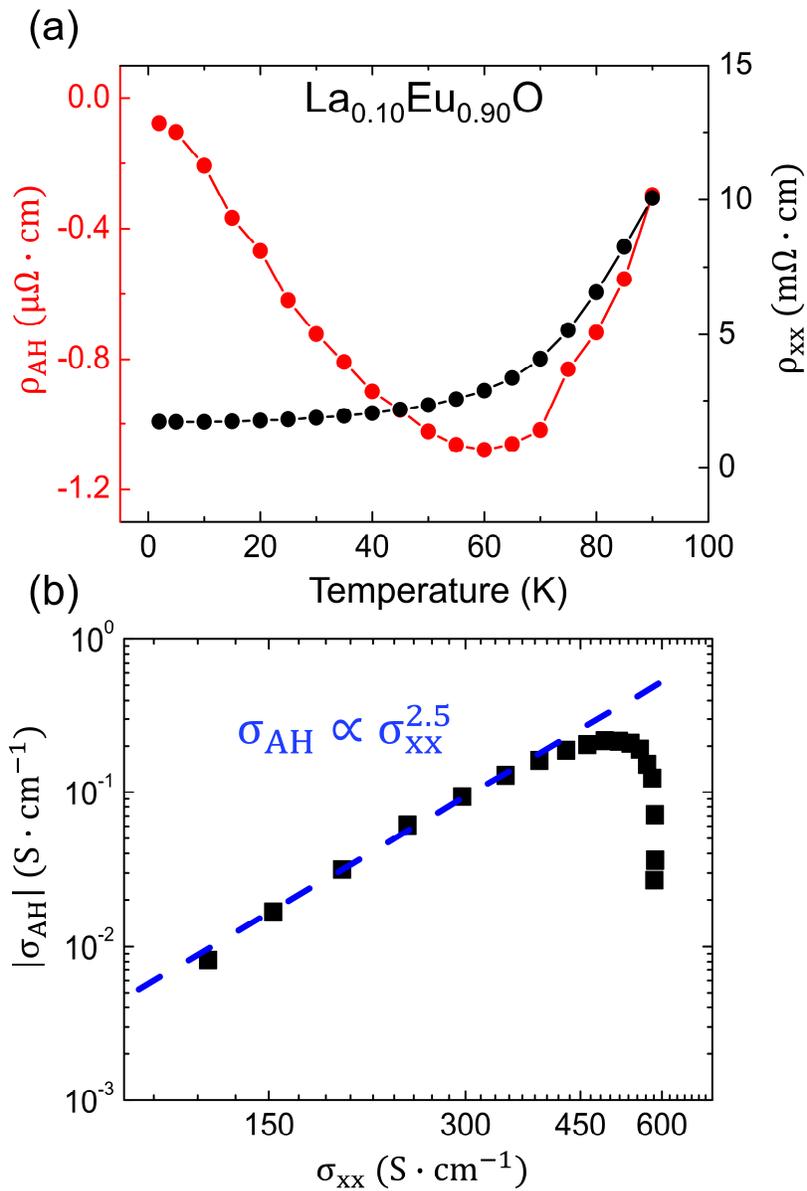

FIG. S2. AHE measurement for the La$_{0.10}$Eu$_{0.90}$O thin film. (a) Anomalous Hall (red) and longitudinal resistivity (black) as a function of temperature. (b) The scaling behavior of anomalous Hall ($\sigma_{AH}$) and longitudinal ($\sigma_{xx}$) conductivities. The blue dashed line represents a relationship of $\sigma_{AH} \propto \sigma_{xx}^{2.5}$.

### S3. Unconventional anomalous Hall effect in La$_x$Eu$_{1-x}$O with low doping

For these La$_x$Eu$_{1-x}$O thin films with low La doping, unconventional anomalous Hall effect are observed at low temperatures, following the similar procedure to obtain the topological Hall effect for high doping samples. The phase diagrams of the unconventional anomalous Hall resistivity ($\rho'_{xy}$) of the La$_{0.005}$Eu$_{0.995}$O and La$_{0.009}$Eu$_{0.991}$O thin films are shown in Fig. S3(a) and S3(b) as a function of the magnetic field and temperature. At low temperatures, the amplitude and sign of the anomalous Hall resistivity exhibit some unconventional features when the external magnetic fields are lower than the saturation fields. The observed unconventional anomalous Hall effect in these two films with low La doping happens at low temperatures, which are totally different from the THE observed in high doping films at higher temperatures due to the formation of skyrmions. To fully understand the unconventional anomalous Hall effect, further studies are needed. This observation also suggests the critical role of the La doping for the skyrmions in the La$_x$Eu$_{1-x}$O thin films.

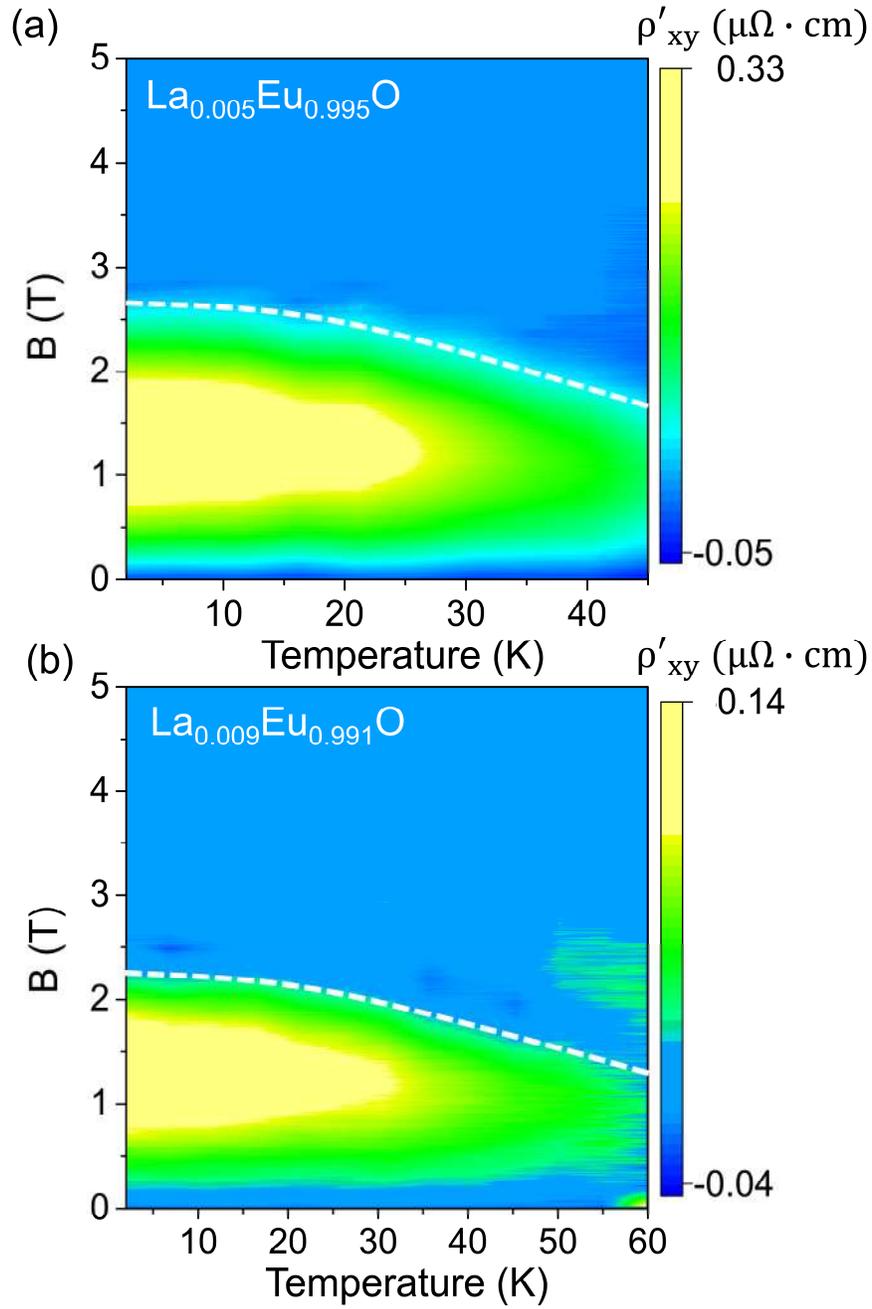

FIG. S3. Unconventional anomalous Hall resistivity for $La_xEu_{1-x}O$ thin films with low doping. (a-b) AHE results in $La_{0.005}Eu_{0.995}O$ and $La_{0.009}Eu_{0.991}O$ thin films. The white dashed lines are guiding lines of the saturation magnetic fields.